\documentclass[aps,prd,showpacs,letterpaper,preprint]{revtex4-1}

\usepackage{graphicx}

\begin{document}

\title{Production of tidal-charged black holes at the Large Hadron Collider}

\author{Douglas M. Gingrich}

\altaffiliation[Also at ]{TRIUMF, Vancouver, BC V6T 2A3 Canada} 

\email{gingrich@ualberta.ca}

\affiliation{Centre for Particle Physics, Department of Physics, University
of Alberta, Edmonton, AB T6G 2G7 Canada}

\date{\today}

\begin{abstract}
Tidal-charged black hole solutions localized on a three-brane in the
five-dimensional gravity scenario of Randall and Sundrum have been known 
for some time.
The solutions have been used to study the decay, and growth, of black 
holes with initial mass of about 10~TeV.
These studies are interesting in that certain black holes, if produced
at the Large Hadron Collider, could live long enough to leave the
detectors. 
I examine the production of tidal-charged black holes at the Large Hadron 
Collider and show that it is very unlikely that they will be produced
during the lifetime of the accelerator.
\end{abstract}


\maketitle

\section{Introduction\label{sec1}}

Models of low-scale gravity could allow particle physics experiments to
study the strong-gravity
regime~\cite{Arkani98,Antoniadis98,Randall99a,Dvali07a,Calmet07}. 
One consequence of these models is the possibility of black hole
production at the Large Hadron Collider (LHC) if the fundamental scale
of gravity $M_D$ is about a
TeV~\cite{Argyres98,Banks99,Dimopoulos01,Giddings01a}.  

Black hole solutions in higher dimensions have a complicated dependence
on both the gravitational field of the brane and the geometry of the
extra dimensions.
The models with large extra dimensions in a flat spacetime
(ADD)~\cite{Arkani98,Antoniadis98} are popular.
One reason for this interest is that if the geometrical scales of the
extra dimensions are all large compared to $1/M_D$, then there is a wide 
regime in which the geometry of the extra dimensions plays no essential
role.  
In ADD this means considering black holes with horizon radius much
smaller than the size of the extra dimensions. 
For large flat compactified extra dimensions, the topology of the black
hole can be assumed to be spherically symmetric in the $(n+3)$-spatial
dimensions and the boundary conditions from the compactification can be
neglected. 
This allows the simple Myers-Perry metric~\cite{Myers} to be used. 
This metric is also useful in the Randall-Sundrum
scenario~\cite{Randall99a} if the warping scale or (AdS radius) is large
compared to $1/M_D$~\cite{Giddings01a}.  
  
In a regime in which the brane can not be neglected, the so called brane
world scenario, there are very few physical solutions to the
higher-dimensional black hole problem (see Ref.~\cite{Gregory08,Kanti09}
for a review).  
One of the earliest and most widely discussed solutions is a black string
solution~\cite{Chamblin99}, but this solution is far from physical.
Actually, it is unknown if physical black hole solutions exist at all in
the Randall-Sundrum model.

A class of Randall-Sundrum solutions have been found in which the
solution is a Reissner-N{\"o}rdstrom metric with the electric charge
replaced by a tidal charge~\cite{Dadhich00}.
If the effects of this tidal-charge term are negligible, the solution
becomes an effective four-dimensional solution and the effects of
low-scale gravity are unlikely to be observed.
If the extra term is significant, the effects of low-scale gravity may
be observable if the fundamental scale of gravity is low enough.

The decays of tidal-charged black holes in the contexts of the
microcanonical picture have been
studied~\cite{Casadio02,Schelpe,Casadio09a,Casadio09b}.
The microcanonical corrections may be significant when the object
reaches the Planck size and the classical black hole description fails 
(For the case of the microcanonical treatment of ADD black hole decays
at the LHC, also see Ref.~\cite{Gingrich07b}.). 
The microcanonical corrections to the canonical decay treatment are
larger in the Randall-Sundrum scenario for Planck-sized objects,
and in certain cases they may live long enough to be considered
quasi-stable.       
The possibility that black holes produced at the LHC might be
quasi-stable has raised safety concerns (see for e.g.\
Ref~\cite{Plaga}), which have been addressed in general
\cite{Giddings08a,Giddings08b,Koch09a}, and in the context of tidal
charged black holes~\cite{Schelpe,Casadio09a,Casadio09b}.       
I will have nothing new to add to the decay discussion.

 
Based on recent phenomenological constraints on the tidal
charge~\cite{Casadio09b}, I estimate the production cross section at LHC
energies and show that it is tiny over most of the allowed parameter
space. 
The probability of producing these black holes during the lifetime
of the LHC accelerator is vanishingly small.
Hence, a scenario of high missing-energy signatures of black holes at
the LHC, because of a long lifetime, is disfavoured.   

\section{Production Cross Section\label{sec2}}

A static solution for a mass $M$ localized on a three-brane in the
five-dimensional gravity scenario of Randall-Sundrum is~\cite{Dadhich00} 

\begin{equation} \label{eq1}
-g_{tt} = (g_{rr})^{-1} = 1 - \frac{2M}{M_p^2} \frac{1}{r} +
\frac{q}{M_5^2} \frac{1}{r^2}\, , 
\end{equation}

\noindent
where $q$ is a dimensionless tidal charge, $M_p = 1.2\times 10^{16}$~TeV
is the effective Planck scale on the brane, and $M_5$ is the fundamental
Planck scale in the five-dimensional bulk. 
Equation~(\ref{eq1}) is an exact solution of the effective Einstein
equations on the brane and represents the induced metric on the brane in
the strong-gravity regime.  
Attempts to extend the metric off the brane have
failed~\cite{Gregory08,Kanti09} so it is hoped that the brane
characteristics of the black hole dominate.  

The tidal term arises due the gravitational affects from the fifth
dimension and the dimensionless parameter $q$ can be thought of as a
tidal charge arising from the projection onto the brane of free
gravitational field effects in the bulk.   
The parameter $q$ is determined by both the brane source, i.e.\ the mass
$M$, and any Coulomb part of the bulk Weyl tensor that survives when $M$
is set to zero~\cite{Dadhich00}.  
It may be positive or negative.
Ref.~\cite{Dadhich00} argues that negative $q$ is more natural.
In this case, there will be one horizon that is outside the Schwarzschild
horizon $R_S = 2M/M_p^2$. 
The bulk effects tend to strengthen the gravitational field.

The tidal charge affects the geodesics and the gravitational
potential, and hence indirect limits may be placed on it by
observations. 
The effects of the tidal term die off quickly with increasing $r$.
At astrophysical distances, the weak-field effects are being probed and
the correction term is much less than the Schwarzschild term in the
effective potential.
However, in the strong-gravity regime $q$ can be large, and the horizon 
radius would become

\begin{equation} \label{eq2}
R_H \sim \frac{\sqrt{-q}}{M_5} \, .
\end{equation}

\noindent
The tidal term will dominate the Schwarzschild term, which reflects the
fact that gravity becomes effectively five-dimensional at high
energies. 
Current observations place only weak limits on the tidal
charge~\cite{Dadhich00}, but in principle significant tidal
modifications could arise in the strong-field regime.
Thus it is useful to investigate the phenomenology of these black holes
at the LHC.

The tidal-charge parameter can only be deduced on dimensional grounds.
Ref.~\cite{Dadhich00} suggests a value of $q = -M/M_{5}$.
Ref.~\cite{Casadio02} generalized this to $q = - (M_p/M_{5})^\alpha
(M/M_{5})$, where $\alpha$ is a real parameter.
For the tidal term to dominate, $\alpha \gg -4$.
Ref.~\cite{Casadio02,Schelpe,Casadio09a} took as a typical choice
$\alpha \ge 0$.
A recent more general parameterization is given by~\cite{Casadio09b}  
 
\begin{equation} \label{eq3}
q = -\left( \frac{M_p}{M_5} \right)^\alpha \left( \frac{M}{M_5}
\right)^\beta,  
\end{equation}

\noindent
where $\beta > 0$ is a real parameter.
Thus the original calculations used the special case of $\alpha = 0$ and
$\beta = 1$.  
For the tidal-charged black hole to be semiclassical, $\alpha \gtrsim
-2$~\cite{Casadio09b}. 

A feature of black hole production in particle collisions is that its
cross section is essentially the horizon area of the forming black hole
and grows with the centre of mass energy of the colliding particles as
some power. 
In the ADD scenario, this cross section is rather large and given by 

\begin{equation} \label{eq4}
\sigma_\mathrm{ADD} \sim \frac{1}{M_D^2} \sim 1~\mathrm{nb}\, ,
\end{equation}

\noindent
where $M_D \sim 1$~TeV is the fundamental Planck scale in $D$ spacetime
dimensions.
For the tidal-charged black hole case,

\begin{equation} \label{eq5}
\sigma \sim \frac{-q}{M_5^2} \sim \frac{1}{M_5^2} \left( \frac{M_p}{M_5}
\right)^\alpha \left( \frac{M}{M_5} \right)^\beta. 
\end{equation}

\noindent
For this cross section to be comparable to the ADD case, $|q|$ must be
of $\mathcal{O}(1)$. 
Recent calculations~\cite{Casadio02,Casadio09a} used $\alpha \ge 0$
and $\beta = 1$, which could give a huge cross section and exceed the
ADD case. 
However, $\alpha $ can be restricted from above by precision
measurements of the deviation of Newton's law~\cite{Kapner07} and thus
$\alpha >0$ may not be a valid choice.

In estimating the upper bound on $\alpha$, I follow a similar argument
to Ref.~\cite{Casadio09b}. 
The brane has a thickness $L$, which must be less than the length over
which deviations from Newton gravity are not observed:
$L \lesssim 44~\mu\mathrm{m} = 2.2\times
10^{14}$~TeV$^{-1}$~\cite{Kapner07}. 
For the tidal-charge term to dominate over the Schwarzschild term,
the horizon radius must be less than the critical radius $r_c$, which
can be defined to be the radius at which the Schwarzschild term equals
the tidal term in the metric:

\begin{equation} \label{eq6}
r_c = \frac{1}{2M_p} \left( \frac{M_p}{M_5} \right)^{3+\alpha} \left(
\frac{M}{M_5} \right)^{\beta-1}.
\end{equation}

\noindent
In addition, this critical radius must be less than the brane thickness,
$r_c \ll L$, else we would have already observed deviations in Newton's
law due to the tidal term.
Thus, requiring $r_c \ll L$ gives $\alpha \lesssim \alpha_c$, where

\begin{equation} \label{eq7}
\alpha_c \equiv \frac{\ln(2L M_p) + (1-\beta)\ln(M/M_5)}{\ln(M_p/M_5)} -
3\, . 
\end{equation}

In low-scale gravity, black holes could be produced at the LHC in the
range $1 \lesssim M \lesssim 14$~TeV.
The upper bound is given by the maximum LHC energy, which would be very
unlikely, and the lower bound is given by the lower experimental limit
on $M_5$. 
For $M_5 = 1$~TeV, $\alpha \lesssim -1.09$ for $M = M_5$, independent of
$\beta$, and $\alpha \lesssim -1.02$ for $M = 14$~TeV, which occurs at
$\beta = 0$. 
For larger values of $M_5$, $\alpha \lesssim -0.94$ for the extreme case
of $M = M_5 = 14$~TeV, which is highly unlikely.
The LHC is a proton-proton collider and thus the probability of
producing a parton-parton collision at the maximum LHC energy is
vanishingly small. 
Because of this, the upper bound on $\alpha_c$ is likely to be less than 
$-0.94$ for realizable values of high $M = M_5$.
Ignoring factors of two in the metric or picking a different definition
for the fundamental Planck scale $M_5$ changes $\alpha_c$ by at most
about 0.1, and in no case is $\alpha_c$ larger than about $-1.0$. 

Assuming the largest allowed value of $\alpha$ is $-1$, gives the upper 
bound on the parton cross section of

\begin{equation} \label{eq8}
\sigma \lesssim \frac{1}{M_5 M_p} \left( \frac{M}{M_5} \right)^\beta
\lesssim \frac{\left(M/\mathrm{TeV}\right)^\beta}{M_p/\mathrm{TeV}}\
\mathrm{TeV}^{-2}\, , 
\end{equation}

\noindent
where in the last expression I have used $M_5 = 1$~TeV, which gives the 
maximum cross section.
Because of $M_p$ in the denominator, the cross section will be
exceedingly low unless $\beta$ is very large.
In terms of numerical values 

\begin{equation} \label{eq9}
\sigma \lesssim 10^{-16+\beta} \sim 10^{-10+\beta}~\mathrm{fb}\, .  
\end{equation}

\begin{figure}[t]
\begin{center}
\includegraphics[width=\columnwidth]{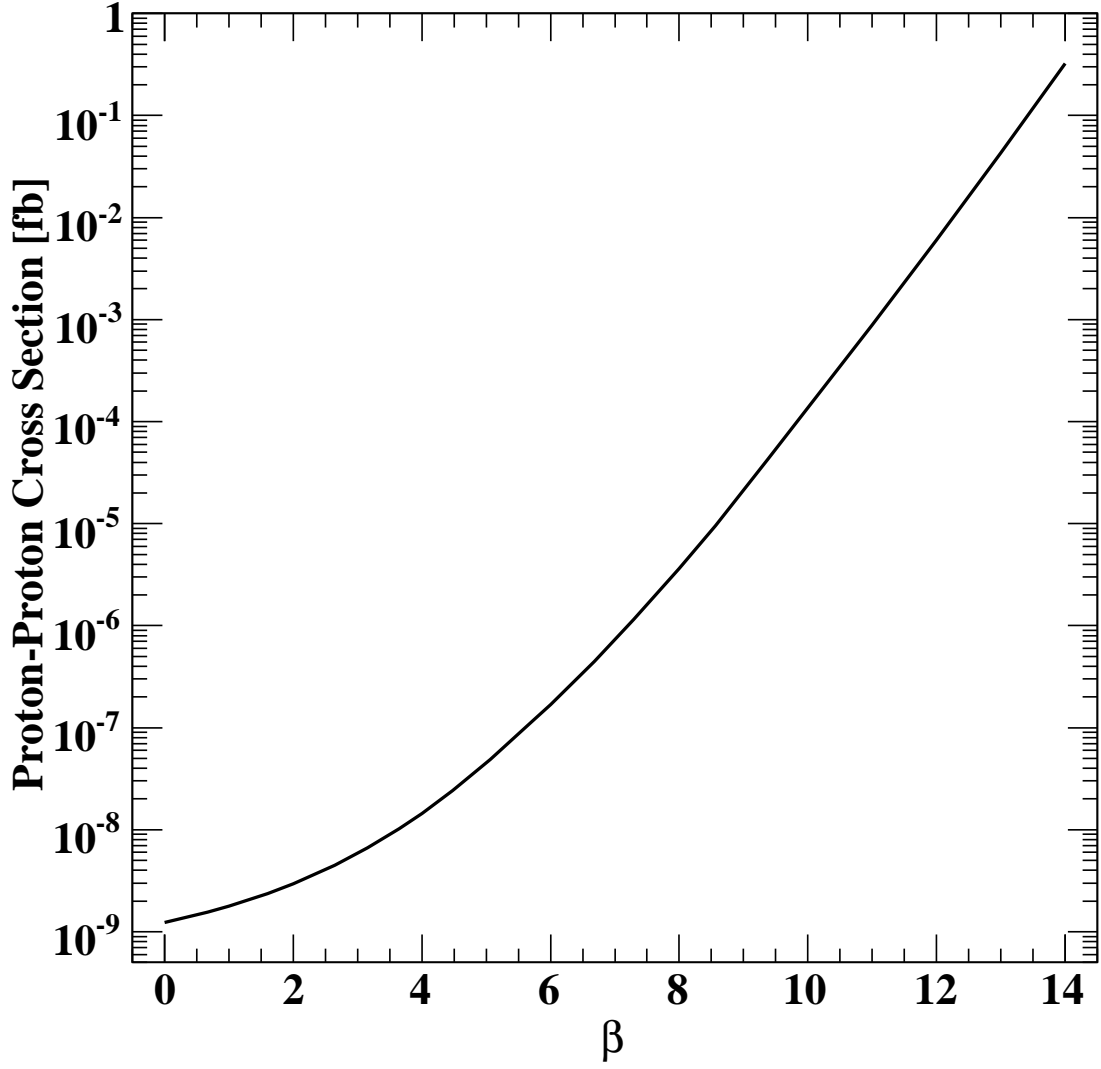}
\caption{Proton-proton cross section versus the $\beta$ parameter for
$M_5 = 1$~TeV.}
\label{fig1}
\end{center}
\end{figure}

To calculate the proton-proton cross section, the parton cross section in
Eq.~(\ref{eq8}) must be convoluted with the parton distribution
functions of the proton, summed over all partons that could form the
black hole, and integrated over all black hole masses.
I will integrate over the full kinematically allowed range although
the applicability of Eq.~(\ref{eq8}) may be questionable near $M \sim 
M_5$.
To ignore this region would significantly reduce the cross section.
The results are shown in Fig.~\ref{fig1} as a function of the $\beta$
parameter for $M_5 = 1$~TeV.

I assume the total integrated luminosity delivered by the LHC will
be 1~ab$^{-1}$.
Thus the probability of producing a single tidal-charged black hole
event during the lifetime of the accelerator is vanishingly small for
reasonable values of $\beta$.
Only for $\beta \gtrsim 11$ will black hole production become
significant.
We would expect $\beta$ to be a few and large values of $\beta$ are
unnatural~\cite{Casadio09b}.  
The functional form for $q$ in Eq~(\ref{eq3}) can not be valid over all
mass scales and some mass dependence of the parameters $\alpha$ and
$\beta$ is expected. 

The $\pi R_H^2$ form for the cross section is usually considered a
maximum cross section.
Possible form factors could increase the cross section by a factor of
about three.
There is far greater discussion on how the cross section is reduced from
the geometrical form.
For example, initial-state beam radiation or radiation during a balding
phase could reduce the cross section by several orders of magnitude (see 
Ref.~\cite{Gingrich06a} for a review of the ADD case). 

The analysis has been confined to the brane and the bulk space has not
been taken into account.  
Some approximations are involved in the terms that encode the bulk
behaviour on the brane.
Assuming these simplifications do not alter the underlying
phenomenology, it appears very unlikely that tidal-charged black holes 
could be produced at the LHC.


\begin{acknowledgments}
This work was supported in part by the Natural Sciences and Engineering
Research Council of Canada.
\end{acknowledgments}

\bibliography{gingrich}
\end{document}